\documentclass[a4paper,11pt]{article}
\pdfoutput=1 

\usepackage{jinstpub} 
\usepackage{verbatimbox}

\title{\boldmath EXFOR-NSR PDF database: a system for nuclear knowledge preservation and data curation}

\author[a]{V.V. Zerkin}
\author[b]{B. Pritychenko,\note{Corresponding author.}}
\author[b]{J. Totans}
\author[a]{L. Vrapcenjak}
\author[c]{A. Rodionov}
\author[c]{G.I. Shulyak}

\affiliation[a]{Nuclear Data Section, International Atomic Energy Agency, \\ Vienna International Centre, P.O. Box 100, A-1400 Vienna, Austria}
\affiliation[b]{National Nuclear Data Center, Brookhaven National Laboratory, \\ Upton, NY 11973-5000, USA}
\affiliation[c]{B.P. Konstantinov Petersburg Nuclear Physics Institute, \\ Orlova Roscha, Gatchina 188300, Leningrad District, Russian Federation}

\emailAdd{pritychenko@bnl.gov}

\abstract{
Current needs of nuclear science and technology include complete, well-documented, and easily verifiable nuclear data. The complete data records require supporting nuclear bibliography, presently stored in dedicated libraries, in addition, to actual data. Experimental nuclear reaction data (EXFOR) and Nuclear Science References (NSR) databases contain compilations based on primary (journals) and secondary (conference proceedings, theses, preprints, etc.) publications, and data received from authors via private communications. The secondary library materials and private communications often represent a bottleneck for nuclear data verification, compilation, evaluation, and dissemination activities. To address this issue, bibliographic materials were scanned into PDF (Portable Document Format) files and uploaded in a relational database.

The traditional scope of nuclear databases that includes meta-data and numbers derived from data in specialized formats was broadened to accommodate the large volumes of original nuclear data publications. The complete PDF publication files were stored in a relational database as Binary Large OBjects (BLOB). This unique collection of nuclear data compilations and supporting publications generate many opportunities for machine learning applications. 

The Web interfaces for authorized and public access to the EXFOR-NSR nuclear publications database were implemented at the U.S. National Nuclear Data Center, {\it  https://www.nndc.bnl.gov} and IAEA Nuclear Data Section, {\it https://www-nds.iaea.org}. The current system is complementary to major nuclear libraries and narrowly focused on nuclear data compilation and evaluation procedures. The contents of the PDF database, details of implementation, and Web interface are described. New capabilities for data curation, knowledge preservation, worldwide dissemination, and natural language processing (NLP)  applications are given. 

}


\begin{document}
\maketitle
\flushbottom

\section{Nuclear Data Compilations and Evaluations}
\label{sec:ND}
Worldwide efforts in nuclear science and technology require the development and aggregation of fully traceable and well-documented data records, including the original data publications and related data sets. These underlying materials are often unique and inaccessible to regular users. Many unique publications are stored as paper copies in specialized nuclear data libraries at the U.S. National Nuclear Data Center (NNDC), Brookhaven National Laboratory (BNL) and Nuclear Data Section (NDS), International Atomic Energy Agency (IAEA) in support of Experimental nuclear reaction data (EXFOR)~\cite{Zer18}, Nuclear Science References (NSR)~\cite{Pr11}, and eXperimental Unevaluated Nuclear Data List (XUNDL)~\cite{xundl}  database compilations. Therefore, it represents an interest to explore the nuclear databases and associated references for the creation of self-contained nuclear data sets.

The EXFOR, XUNDL, and NSR compilations create a basis for the Evaluated Nuclear Data File (ENDF)~\cite{endf,Chadwick11} and Evaluated Nuclear Structure Data File (ENSDF)~\cite{ensdf} evaluations produced by evaluators around the globe. Nuclear data compilations and evaluations activities are managed in the USA by the United States Nuclear Data Program (USNDP)~\cite{usndp,Ber20,Pri20} and the Cross Section Evaluation Working Group (CSEWG)~\cite{CSEWG}. International data efforts are led by the Nuclear Data Section, IAEA~\cite{IAEA} and Organisation for Economic Co-operation and Development (OECD), NEA-Data Bank~\cite{OECD}. The USNDP provides technical expertise and resources for the NSR database. The IAEA, in collaboration with the USNDP and other member states, provides support for the Nuclear Reaction Data Centers (NRDC)~\cite{NRDC,Otuka14} and the Nuclear Structure and Decay Data (NSDD)~\cite{NSDD,Dim20} networks that are responsible for the present-day operations of the EXFOR and XUNDL/ENSDF databases, respectively.

\subsection{Compilation and Evaluation Workflow}
\label{sec:workflow}

Nuclear data play a critical role in nuclear energy production, national security applications, and computer code developments. The quality of input numerical data assures confidence in calculated results. For this reason, the nuclear data application developers rely on thoroughly prepared evaluated nuclear data sets. The evaluated data sets are based on numerical values in experimental nuclear reaction, structure, and decay compilations created by data scientists at the major national laboratories and universities where the nuclear data are produced. Scientific research funding organizations stand firm on the meticulous verification of compiled values against the published numbers and comprehensive data sets~\cite{nsac,ostp}. These numbers and data sets are obtained using the hosting institutions' journal subscriptions, lab reports, conference proceedings, and interactions with the data producers. Over the last 70 years, many of these priceless data sets and documents were archived in dedicated libraries at Brookhaven and Vienna.

The data professionals always sought access to the Brookhaven and IAEA library resources used in the EXFOR, XUNDL, and NSR compilations, and satisfying the individual requests became a time-consuming activity for both organizations. The contemporary Web and relational database management systems (RDBMS) progression provided an opportunity for resolving this problem. Approximately $80\%$ of original references were collected as PDF files and $\sim$ 220,000 files were stored in a relational database.  
The ongoing scanning effort at Brookhaven Lab aims to provide complete coverage of rare Ph.D. Theses, conference proceedings, laboratory reports, and private communications. The above-mentioned scanning activities helped to manage the nuclear data evaluations workflow in the domestic and international networks, replace the bulk of traditional nuclear physics libraries with their electronic counterparts and improve users' experiences. 

The EXFOR-NSR PDF database is not uncommon. The value of a systematic collection of nuclear publications was recognized by the International Atomic Energy Agency, and the International Nuclear Information System (INIS) was launched by 1970~\cite{inis,Mor70}. The INIS membership consists of 132 countries and 24 international organizations, and the project scope and database keywords are very broad. Over the years, they accumulated a diverse collection of published materials in all areas related to peaceful uses of nuclear science and technology~\cite{Tol10,Sav13}. The international bibliography system includes over four million bibliographical records and 600,000 full-text documents. This unique bibliography collection is a treasure trove of nuclear information that requires extensive database knowledge and advanced user skills. In the United States, 150,000 volumes and 1.5 million unclassified nuclear physics and engineering reports are assembled at the Los Alamos National Laboratory (LANL) research library. The LANL library grants full access to its reports from 1943 until 2005 via the Primo search tool~\cite{Primo} while other library resources are accessible to local users only.  These databases have a much larger compilation scope and size compared to the highly-specialized EXFOR-NSR database.  Meanwhile, nuclear data compilers, evaluators, and scientists often need a smaller nuclear bibliography system that is customized for their needs. The constructive interactions on nuclear bibliography between the authors, data evaluators, and user groups led to the creation of the PDF database using the EXFOR and NSR databases contents which will be further discussed in the next subsections.

\subsection{EXFOR Database}
\label{sec:X4}

The EXchange FORmat (EXFOR) library~\cite{Zer18} was created in support of nuclear energy applications, fundamental research, and evaluation activities. 
The initial database scope was limited to neutron-induced reaction quantities. Later, the scope was extended to include charged-particle and photon-induced reactions with energies below the pion production threshold.  Presently, neutron-, proton-, alpha-, and photon-induced reactions constitute 47.9, 19.8, 7.36, and 6.2 $\%$ of database contents, respectively.  
In addition to cross-section data, the EXFOR library includes information on particle spectra such as $^{252}$Cf spontaneous fission and $\beta$-delayed neutrons. 
Heavy-ion-, electron- and pion-induced reactions with energies up to 1 GeV are compiled by the responsible centers voluntarily. 
The EXFOR database is the largest low- and intermediate-energy nuclear reaction library; as of October 14, 2021, it includes 23,887 experiments, 
177,447 data sets, and 18,876,875 data points. The database contains information on 1,121 targets, 485 incident projectiles, and 2,759 nuclear reactions. Individual database compilations (experiments) were assembled using all relevant references,  $\sim$32,000 in total.  
It is well known that nuclear reaction measurements are often unique and expensive. A recent analysis of required fundings for a new experiment showed the $\$$1 M price tag~\cite{20Pri}. The compiled data are the treasure trove of information and a supporting nuclear bibliography is needed for data sets completeness and verification purposes.

\subsubsection{EXFOR Data Compilation}

Organized nuclear data activities originated from the Manhattan Project~\cite{Gold47,Wil47,Chad21}. In the subsequent years, many Manhattan Project alumni migrated to a newly created Brookhaven National Laboratory to continue the original work.  In the early 1950s, data compilations were well organized at Brookhaven in support of nuclear science and reactor research activities.  Since 1964 Brookhaven compilations have been stored in the Sigma Center Information Storage and Retrieval System (SCISRS) that predated the EXFOR database.  These experimental data compilation efforts have always had an important international component. The IAEA Nuclear Data Section has been involved in this work since its creation in 1964. Other early contributors include NEA Data Bank, Paris, France, and the Institute of Physics and Power Engineering, 
Obninsk, USSR which were founded in 1964 and 1963, respectively \cite{Zer18}. In 1969 an agreement on an exchange format was reached between four centers and July 1970 was chosen as the starting date for transmission of neutron data among the participating centers in the EXFOR data interchange format. The pre-1976 compilation scope of neutron cross sections and spontaneous fissions was defined by the needs of an ENDF project. An example of neutron cross section compilation is shown in Fig.\ref{fig:x4comp}. 
\begin{figure}[ht!]
\begin{verbnobox}[\fontsize{8pt}{8pt}\selectfont]

ENTRY            14511   20181206   20190405   20190318       1446
SUBENT        14511001   20181206   20190405   20190318       1446
BIB                  9         12
TITLE      Cross Sections for the Reactions C12(p,pn)(n,2n)C11
AUTHOR     (S.D.Warshaw,R.A.Swanson,A.H.Rosenfeld)
REFERENCE  (J,PR,95,649(SA2),1954)
REL-REF    (O,C2359001,S.D.Warshaw+,J,PR,95,649(SA2),1954)
           C12(p,pn) data.
INSTITUTE  (1USACHI)
FACILITY   (SYNCY,1USACHI) The University of Chicago
           synchrocyclotron.
METHOD     (ACTIV)
ERR-ANALYS (DATA-ERR) Uncertainty in calibration of the C11
           counter.
HISTORY    (20181206C) BP
ENDBIB              12
NOCOMMON             0          0
ENDSUBENT           15
SUBENT        14511002   20181206   20190405   20190318       1446
BIB                  2          2
REACTION   (6-C-12(N,2N)6-C-11,,SIG)
STATUS     (TABLE) page 649.
ENDBIB               2
NOCOMMON             0          0
DATA                 3          1
EN-APRX    DATA       DATA-ERR
MEV        MB         MB
      400.0       17.9        1.4
ENDDATA              3
ENDSUBENT           10
ENDENTRY             2

\end{verbnobox}
\caption{EXFOR database record in the reaction exchange format includes a unique identifier or accession number (14511), bibliographical information, keywords, record history, and cross section data.}
\label{fig:x4comp}
\end{figure}

Later on, the EXFOR compilations became even more popular worldwide, and many institutions have joined. The database compilations represent one of the oldest continuously operated scientific collaborations. Since its creation, the EXFOR project has relied heavily on computer technologies available at the time. Over the years, the reaction data compilations have evolved from a pencil and paper operation into a technology enterprise that employs relational database and Web servers~\cite{pr06}, EXFOR compilation editors~\cite{Pik20}, plot digitizers~\cite{Ot20}, and optical character recognition technologies. 

\subsection{NSR Database}
\label{sec:NSR}
The nuclear structure references (NSR) database~\cite{Pr11} was created at Oak Ridge National Laboratory around 1960 in support of ENSDF evaluations. It is a bibliography of nuclear physics articles, indexed according to content and spanning more than 120 years of research. Over 80 journals are checked every week for articles to be included. Fig.\ref{fig:key} shows an example of a database entry (NSR data record) which includes an NSR keynumber (unique record identifier), bibliographical metadata (authors, title, reference), and keywords in the bibliography exchange and data compilation format. In 1980 the database management was transferred to NNDC, Brookhaven National Laboratory. In the 90s,  the database scope evolved from nuclear structure to general nuclear science, and keywords were broadened to reflect the new project scope. In subsequent years, digital object identifiers (DOI) were added to NSR to accommodate the current publication trends, and extensive coverage of nuclear reactions was included. The NSR compilation policies require unique keywords that reflect new results only, any mentions of previous findings are ignored by the database compilers. 

\begin{figure}[ht!]
\begin{verbnobox}[\fontsize{8pt}{8pt}\selectfont]
<KEYNO   >2010GA14
<HISTORY >A20100806 M20100818
<CODEN   >JOUR PRVCA 81 064326
<REFRENCE>Phys.Rev. C 81, 064326 (2010)
<AUTHORS >A.Gade, T.Baugher, D.Bazin, B.A.Brown, C.M.Campbell, T.Glasmacher, 
G.F.Grinyer, M.Honma, S.McDaniel, R.Meharchand, T.Otsuka, A.Ratkiewicz, J.A.Tostevin, 
K.A.Walsh, D.Weisshaar
<TITLE   >Collectivity at N=50: {+82}Ge and {+84}Se
<KEYWORDS>NUCLEAR REACTIONS {+197}Au({+82}Ge,{+82}Ge'),E=89.4 MeV/nucleon; 
{+197}Au({+84}Se,{+84}Se'),E=95.4 MeV/nucleon; {+9}Be({+82}Ge,{+82}Ge'),E=87.6 MeV/nucleon; 
{+9}Be({+84}Se,{+84}Se'),E=92 MeV/nucleon,[{+82}Ge and {+84}Se secondary 
beams from {+9}Be({+86}Kr,X),E=140 MeV/nucleon]; measured E|g,I|g,|s,(particle)|
g-coin; {+82}Ge,{+84}Se; deduced levels,J,B(E2), T{-1/2}.   Intermediate energy 
Coulomb excitation and inelastic scattering. Comparison with systematics of B(E2
) values for first 2+ state in N=50 isotones for Z(even)=30-42 and even-even Ge 
(A=64-82) and Se (A=68-84) isotopes, and with shell-model calculations. Systematics 
of first 3- states in even-even Se (A=74-82) and N=50 isotones.
<DOI     >10.1103/PhysRevC.81.064326
\end{verbnobox}
\caption{NSR database record in the bibliography exchange format includes a unique identifier or NSR keynumber (2010GA14), record history, metadata, keywords, and doi.}
\label{fig:key}
\end{figure}

 NSR is essential in nuclear data evaluations. For instance, Atomic Mass Evaluation (AME) 2020 / NUBASE 2020~\cite{AME20,Nub20} and Evaluated Nuclear Structure Data File~\cite{ensdf} include 6,137 and 59,093 references, respectively. It is impossible to manage $\sim$65,000 references and author lists without a database. 
As of October 14, 2021, the NSR database~\cite{Pr11} had 240,594 entries and 195,478 keyworded abstracts. The NSR keywords are used by the evaluators to quickly identify the relevant publications; however, the lack of direct library access severely hampered the speed of nuclear data evaluations over the many years.  

The NSR database records are based on collections of journals, books, conference proceedings, theses, laboratory reports, and private communications that were stored in the Oak Ridge and Brookhaven libraries as paper records. In recent years, the Oak Ridge National Laboratory library secondary publications were transferred to Brookhaven and assembled at NNDC. Furthermore, the NNDC library collected many bibliographical papers and microfiche records from the Los Alamos National Laboratory, Texas A$\&$M and McMaster Universities, and individual donations. The above-mentioned collections transformed it into the prime nuclear data library worldwide. To share the library resources with nuclear data networks, electronic access to the EXFOR and NSR publications was developed as described in the next sections.

\subsection{ Experimental Unevaluated Nuclear Data Database}
\label{sec:XUNDL}
The Experimental Unevaluated Nuclear Data (XUNDL) database~\cite{xundl} contains compiled experimental nuclear structure and decay data sets from more than 3,500 recent papers. These compilations are based on NSR bibliography~\cite{Pr11} and used in ENSDF library evaluations~\cite{ensdf}.

\section{Data Storage and Dissemination  System}

Rapid access to nuclear data is of paramount importance for science and technology professionals, and, since 2000, NNDC and NDS invested heavily into the latest Web and database  developments~\cite{pr06}.   
The modern relational database servers allow access to data using a Structured Query Language (SQL). 
The complementary software parses Web requests into corresponding SQL statements that are passed to the database to harvest data.  
Over the years, commercial and license considerations compelled both centers to settle on Apache Tomcat Web and MariaDB servers. 

In subsequent subsections, the authors will describe the contemporary computer hardware and software computer environments, followed by software application developments, careful analysis of Web features, and their worldwide impact on nuclear data usage.

\subsection{PDF Database}

The EXFOR-NSR PDF database consists of two bibliographic files collections, and the database is implemented in the EXFOR relational database schema~\cite{Zer18}. The schema describes a database structure in a formal language supported by the database management system. In a relational database, the schema defines the tables, fields, relationships, views, indexes, procedures, and other elements.  
The EXFOR schema includes tables with information extracted from the EXFOR files (data, metadata, and dictionaries) and relationships between data tables, e.g. ENTRY and REFERS metadata tables are linked via EntryID (relationship one-to-many), although tables can be ``joined" on the fly in a SQL command SELECT statement.  The EXFOR database contents were previously described in Ref.~\cite{Zer18}, here we would concentrate on the PDF database only.

\begin{figure}[ht]
 \centering
\includegraphics[width=0.8\textwidth]{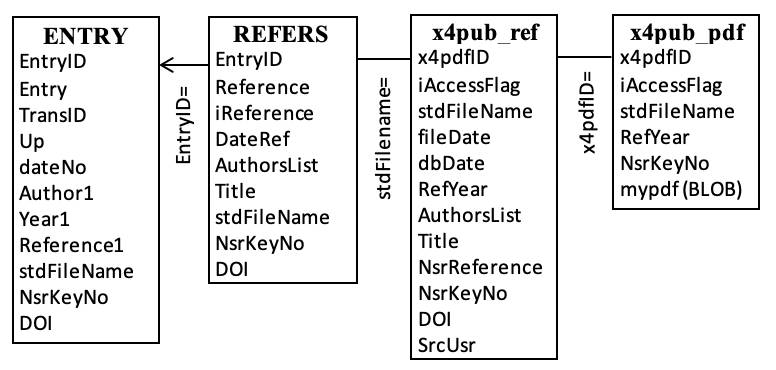}
\caption{The EXFOR-NSR PDF database schema. }
\label{fig:4tables}
\end{figure}

\begin{figure}[ht]
 \centering
\includegraphics[width=0.8\textwidth]{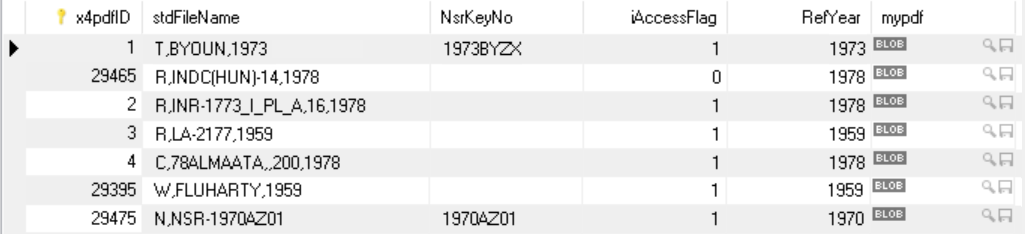}
\caption{The extended view of x4pub$\_$pdf data table: PDF files are stored in the mypdf column as BLOBs. }
\label{fig:x4blob}
\end{figure}

The simplified EXFOR-NSR PDF database schema is shown in Fig.~\ref{fig:4tables}. Two EXFOR database tables ENTRY and REFERS are associated with PDF files stored in  x4pub$\_$pdf, x4pub$\_$ref data tables using join statements. Fig.~\ref{fig:x4blob}  show that x4pub$\_$pdf table contains PDF files while x4pub$\_$ref contains their metadata. Original PDF files are stored in the table x4pub$\_$pdf (column mypdf) as Binary Large OBjects; these files are encrypted to avoid access by unauthorized software. Value in the column iAccessFlag regulates access rights for different categories of users (authorized users and public access). Every PDF file is identified by x4pdfID and described by a unique code stdFileName (generalized code based on EXFOR coding rules for references and EXFOR Dictionaries 5, 6, 7~\cite{X4Dict}) and NSR unique identifier NsrKeyNo. The tables can be easily linked with both EXFOR and NSR databases and various applications software  (such as Web-CINDA, MyEnsdf, IBANDL~\cite{Zer18}).

Comparison of bibliographical information from EXFOR and NSR reveals bibliography coverage overlap and multiple cases when EXFOR references are missing in NSR. Published experimental data absent in EXFOR were extensively discussed in Refs.~\cite{Zer18,20Pri}, while $\sim$22,000 overlapping NSR references supply $\sim$1,400 missing bibliography files for EXFOR database compilations.  Reciprocally, the EXFOR collection provides  $\sim$1,900 additional files for NSR database compilations.  These PDF collections are complementary, {\it i.e.} one collection supplies the missing publications of the other, and vice versa as shown in Table~\ref{Table1}. Further analysis of the Table~\ref{Table1} data shows that the present-day PDF coverage for EXFOR and NSR references are 75.4 and 79.1 $\%$, respectively. 
\begin{table}
\caption{\label{Table1} PDF coverage for EXFOR and NSR references as of October 14, 2021. Reciprocal PDF contributions are shown as \# of complementary files.}
\begin{center}
\begin{tabular}{l|c|c|c}
\hline \hline
Database & \# of References & \# of PDF Files &  \# of Complementary Files \\ 
\hline
EXFOR & 34,609 & 26,343 & 1,899 \\
NSR       & 236,583   &   187,617 & 1,375 \\
\hline \hline
\end{tabular}
\end{center}
\end{table}

\begin{figure}[ht]
 \centering
 \includegraphics[width=0.8\textwidth]{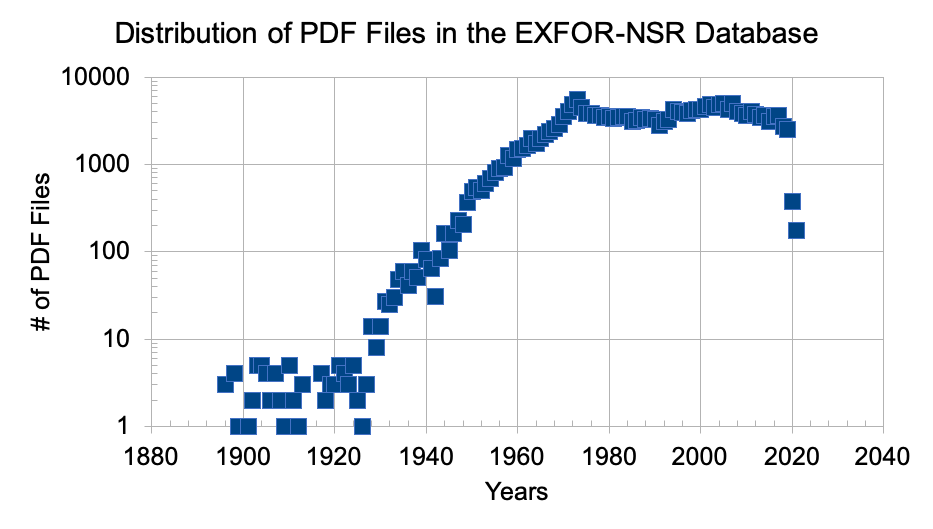}
\caption{Annual distribution of PDF files in the EXFOR-NSR publications database as of October 14, 2021.}
\label{fig:PDFCoverage}
 \end{figure}

Both PDF collections were merged in the present work. The merger is beneficial for nuclear data users because it provides more comprehensive coverage of electronic materials (publications) that were used to produce EXFOR and NSR compilations. The annual distribution of PDF database contents from 1896 to 2021 is shown in Fig.~\ref{fig:PDFCoverage}. PDF contents distribution reflects publication trends in nuclear science over the last 120 years: \# of PDF files is increasing from single digits at the beginning of 20$^{th}$ century until it reaches the present-day level in the 70s. The lack of nuclear physics publications at the beginning of World War I (1914-1916) and a sharp decline during World War II (1942) are visible. Relatively low numbers for 2020 and 2021 reflect the ongoing data collection and publication scanning activities.

\begin{figure}[ht]
\centering
\includegraphics[width=0.9\textwidth]{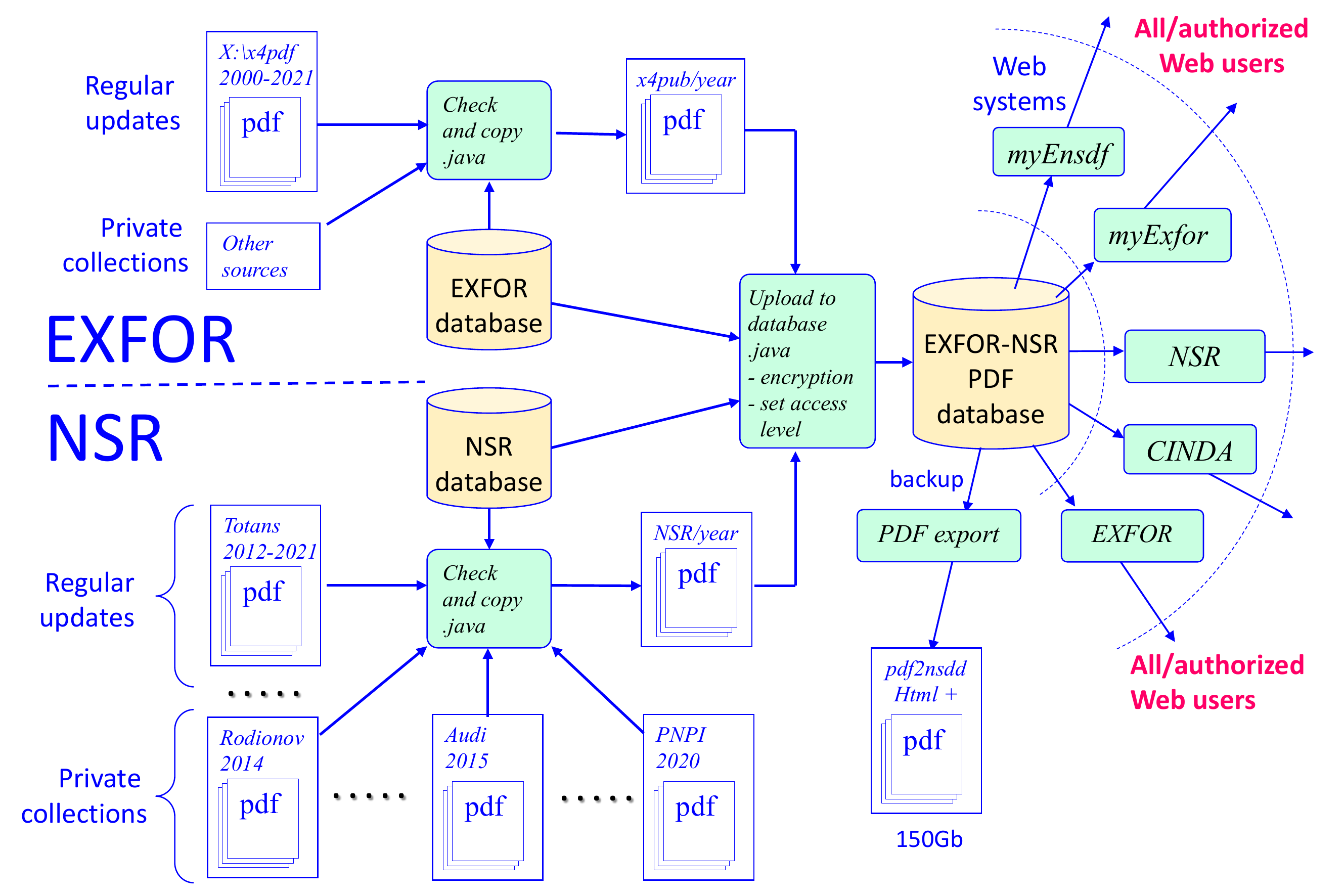}
\caption{PDF database operation management. The PDF database incorporates contributions from EXFOR and NSR, and provides authorized access to nuclear data network members.}
\label{fig:PDFW}
\end{figure}

The overall PDF database operation management is shown in Fig.~\ref{fig:PDFW}. It includes contributions from nuclear reaction, structure, and decay data communities. 
The IAEA NRDC network~\cite{NRDC} supplies the database with new and revised reaction data compilations, dictionaries, and manual updates. Bibliographic metadata, original publication PDF files, and data renormalization information are provided through separate channels. This nuclear science and bibliographical data are processed and stored in a relational database that incorporates text compilation and corresponding PDF files for the majority of EXFOR entries. The NSR operation broadens the scope of database contributions from nuclear reactions to the whole nuclear science, incorporates nuclear data users' comments, and helps with rare publications.   

The total volume of PDF database of $\sim$220,000 electronic files require $\sim$190 GB of x4pub$\_$pdf of data table disk space, and the high-performance soft and hardware are needed for efficient database operations. To satisfy these requirements, a computer configuration that includes two databases, two Web and one development servers was deployed at the Information Technology Department of Brookhaven National Laboratory. The BNL selected Dell PowerEdge R630 servers with Intel Xeon 2.4-GHz processor, 10 cores/processor, 192 GB RAM, disk storage of 5.4 TB (6 units of 15k-RPM, SAS, 512n 2.5-in. Hard Drives), and Red Hat Enterprise Linux v7.9 operating system. Similar computer arrangements were made at the NDS, IAEA.

\section{Current Usage of PDF Database}

For more than 50 years, the two international networks, NRDC~\cite{NRDC} and NSDD~\cite{NSDD}  regularly oversee nuclear reaction, structure, and decay data publications worldwide.  Due to major publishers' subscription rules, the authorized PDF database was implemented. There are several types of product usage:
\begin{itemize}
\item NNDC, BNL, and NDS, IAEA data scientists access the database on the institutions' campuses,
\item Some of the NRDC and NSDD networks members utilize authorized access by supplying their credentials, 
\item Outside users view a small public portion of the database ($\sim$1.2$\%$),
\item All users use doi links to access journal articles from major publishers.
\end{itemize}
If nuclear data compilation and evaluation criteria are satisfied, the PDF file link is generated by the NSR or EXFOR Web Interfaces, and users access the underlying file. The PDF database does not have a dedicated Web Interface; its contents are available within EXFOR and NSR Web applications as supplementary materials. 

The bibliographic materials are crucial in a nuclear reaction, structure, and decay data evaluation work~\cite{Atlas,EnsMan}. The nuclear data curation cycle is complex, and it takes from six months to a few years to complete:
\begin{itemize}
 \item Evaluators study the topic and underlying nuclear physics, 
 \item Search relevant data in EXFOR, NSR, and  XUNDL compilations,  
 \item Access published materials to verify the compiled data,  
 \item Search nuclear databases, Internet, and contact researches for additional information,
 \item Perform nuclear model calculations,
 \item Analyze data, introduce corrections and renormalizations if needed, 
 \item Deduce recommended values,
 \item Validate the results.
 \end{itemize}
The volume of novel experimental data for a particular ENDF target material or ENSDF mass chain justifies launching new or updating the existing evaluations. The new data are discovered through extensive literature and database searches, data community knowledge management, and interactions with external researchers.  Evaluators browse through hundreds of relevant research articles and experimental data sets in NSR and EXFOR databases, respectively, read all applicable publications from the PDF database, investigate data corrections, share important findings, and collect feedback from the data users to improve the recommended values. Attention to detail is required in such work, and rapid access to all material PDF files is essential for timely completion. 

\begin{figure}[]
 \centering
\includegraphics[width=0.8\textwidth]{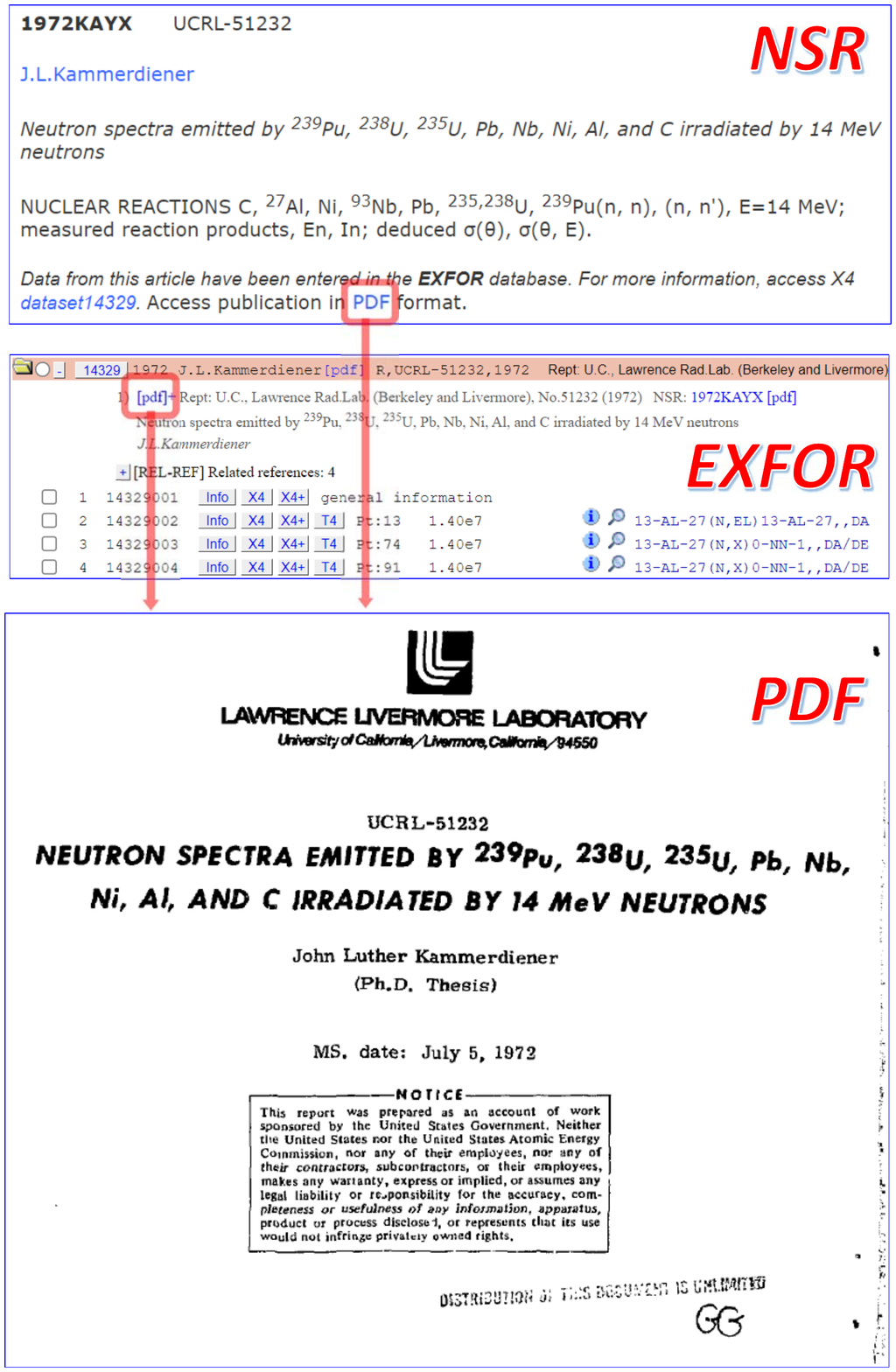}
\caption{EXFOR-NSR retrieval of a J.L. Kammerdiener thesis~\cite{Kam72} PDF file.}
\label{fig:login}
\end{figure}

The PDF Web retrieval provides complementary details on EXFOR compilations and access to numerical values behind the NSR database for nuclear reaction and structure $\&$ decay data evaluators, respectively. An example of the J.L. Kammerdiener thesis~\cite{Kam72} data retrieval that was used to produce the NSR keyworded abstract 1972KAYX and EXFOR compilation $\#$14329 is shown in Fig.~\ref{fig:login}. Kammerdiener thesis compilations are vital for the MCNP code validation and ENDF evaluation efforts because of a large number of unique measurements for a variety of target materials. It was very common 50 years ago to publish important results in Ph.D. theses or laboratory reports without subsequent submission to research journals~\cite{Pri16}. 

The PDF files coverage for relevant data evaluation and computer code validation materials is shown in Table~\ref{Table2}. Table data show the higher coverage of NSR publications compared to EXFOR references. The ongoing scanning of Brookhaven library would address PDF coverage issues for the secondary references in the next few years.
\begin{table}
\caption{\label{Table2} PDF files coverage for several types of NSR and EXFOR references as of October 14, 2021.}
\begin{center}
\begin{tabular}{l|c|c}
\hline \hline
Reference Type & NSR: \#PDF/\#Total &  EXFOR: \#PDF/\#Total \\ 
\hline
Reports	& 14,761/27,895 (53\%) &	2,107/5,397 (39\%) \\
Conf. Proceedings &	13,352/20,143 (66\%)   &	2,031/2,731 (74\%) \\
Theses &	654/2,051 (32\%)   &	100/434 (23\%)  \\
Books &	90/155 (58\%)	& 34/102 (33\%)  \\
Priv. Communications &	1,291/2,107 (61\%)	& 1/815 (0.1\%)  \\
\hline \hline
\end{tabular}
\end{center}
\end{table}
The extensive collections of secondary references coupled with the major nuclear databases make the PDF database an essential data component worldwide, and the NNDC  as well as  Atmospheric Radiation Measurement Data Center, Joint Genome Institute, Materials Project, Particle Data Group, and Systems Biology Knowledgebase (KBase) facilities were granted a status of the SC Public Reusable Research (PuRe) Data Resource~\cite{Jank21,SC21} by the U.S. Department of Energy, Office of Science (SC).

\section{Machine Learning Development using EXFOR-NSR PDF Database}

Thus, the PDF database supplies supporting materials for nuclear physics research, data compilations, and evaluations. The present-day EXFOR and NSR compilations are categorized largely through manual processes by multiple contributors around the world. The $\sim$24,000 EXFOR and $\sim$240,000 NSR compilations can represent very extensive collections of training/testing samples and machine learning (ML)  benchmarks providing a fruitful training ground for the implementation of computer automation algorithms and techniques. The ML natural language processing (NLP) algorithms can read the PDF database contents, generate data outputs, evolve, and improve automatically through experience with the humanly-produced data. An example of an application based on natural language processing is an ongoing project of the Brookhaven-Berkeley-Stony Brook (BBSB) group~\cite{Gem21} for NSR metadata and keywords extraction and data compilations presented below.

The BBSB group intends to use the PDF database for automation of the process of adding new articles to the NSR database by using NLP, development of machine learning algorithms for the identification and application of keywords, expansion of the compilation scope, and adaption of an applied physics-oriented keywords lexicon. The automated keywords will be compared in terms of their ability to precisely capture the semantic content and suitability for USNDP needs. The output of these algorithms will be further explored with human-derived keywords from existent NSR entries for verification and validation of the ML approaches. The project should result in a natural language processing suite optimized for nuclear physics.


\section{Conclusion and Outlook}
\label{sec:Conclusions}

The creation of the EXFOR-NSR PDF database is an important step in the development of complete, well-documented, and easily verifiable nuclear data records. The database is customized for nuclear data activities with a focus on secondary publications.  It is based on unique NNDC, BNL, and NDS, IAEA library resources, provides access to the NSR and EXFOR database bibliographies and simplifies the nuclear data curation workflow worldwide. Due to copyrights restrictions, the database access is restricted to BNL and IAEA scientists and the several NSDD and NRDC international data network participants. There are possibilities for granting public access to selected  PDF files per agreements with individual research organizations: NDS, IAEA, and the Institute for Nuclear Research of the National Academy of Sciences of Ukraine permitted public access to their reports and preprints. NDS, IAEA, and NNDC will continue to address the applied and fundamental science user needs by exploring data access capabilities for laboratory reports per agreements with the individual institutions. These developments would help to increase the PDF database public access coverage from the current $\sim$1.2$\%$ to a higher number and strictly follow the copyright law.

The very promising initiative of the U.S. Department of Energy Office of Scientific and Technical Information (OSTI) on the increase of public access to unclassified scholarly publications and digital data resulting from federal research and development  funding~\cite{OSTI} may result in partial availability of the PDF database for external users. 

Access to the source of EXFOR and NSR data, PDF database gives the possibility not only to validate numerical data and obtain details of experiments but also (a) extract useful information missing in traditional compilations using modern text/image recognition techniques for building curated databases and (b) to build machine learning systems studying various aspects of data to extend and improve evaluation methods by artificial intelligence (AI) technologies. 

The PDF database contributed to the designation of NNDC facilities by the U.S. Department of Energy, Office of Science (SC) as the SC Public Reusable Research (PuRe) Data Resource~\cite{Jank21,SC21}.  The nuclear bibliography database and its Web interface represent a robust and modern system that has evolved over the last 10-15 years of operation. The scalable structure of the PDF database allows the incorporation of other nuclear bibliographic electronic resources (or \textbf{e}-resources) in the future. This bibliographic system is based on the latest computer technologies and aims to satisfy the present and future nuclear data compilers and evaluators' needs, for a broader audience the INIS system and LANL Primo search tool are recommended.

\acknowledgments


The authors are indebted to Alejandro Sonzogni (BNL) and Paraskevi Dimitriou (IAEA) for support of this project, David Brown, Ramon Arcilla (BNL) for useful comments, Svetlana Dunaeva (IAEA), Balraj Singh (McMaster University), Filip Kondev (Argonne National Laboratory), Georges Audi (CSNSM, IN2P3-CNRS), the NRDC and NSDD network members for individual contributions of rare documents. 
Work at Brookhaven was funded by the Office of Nuclear Physics, Office of Science of the U.S. Department of Energy, under Contract No. DE-SC0012704 with Brookhaven Science Associates, LLC.  \\ \\


\end{document}